\documentclass{article}
\usepackage{amsmath}
\usepackage[dvips]{graphics}
\usepackage{latexsym}

\begin{document}
\newcommand{\bea}{\begin{eqnarray*}}
\newcommand{\eea}{\end{eqnarray*}}
\newcommand{\bean}{\begin{eqnarray}}
\newcommand{\eean}{\end{eqnarray}}
\newcommand{\eqs}[1]{Eqs. (\ref{#1})}
\newcommand{\eq}[1]{Eq. (\ref{#1})}
\newcommand{\meq}[1]{(\ref{#1})}
\newcommand{\fig}[1]{Fig. \ref{#1}}

\newcommand{\tri}{\delta}
\newcommand{\grad}{\nabla}
\newcommand{\pa}{\partial}
\newcommand{\pf}[2]{\frac{\pa #1}{\pa #2}}
\newcommand{\cla}{{\cal A}}
\newcommand{\aqt}{\frac{1}{4}\theta}

\newcommand{\om}{\omega}
\newcommand{\omo}{\omega_0}
\newcommand{\mo}{ M_{\rm o}}
\newcommand{\mi}{ M_{\rm i}}
\newcommand{\qo}{ Q_{\rm o}}
\newcommand{\qi}{ Q_{\rm i}}
\newcommand{\rop}{r_{{\rm o}+}}
\newcommand{\rom}{r_{{\rm o}-}}
\newcommand{\rip}{r_{{\rm i}+}}
\newcommand{\rim}{r_{{\rm i}-}}

\newcommand{\oh}{\frac{1}{2}}
\newcommand{\hsp}{\hspace{0.1mm}}
\newcommand{\spa}{\hspace{3mm}}
\newcommand{\hst}{\ \ \ }
\newcommand{\upd}[2]{^#1\hsp_#2}
\newcommand{\ri}{R_i}
\newcommand{\ro}{R_o}
\newcommand{\eqn}{&=&}
\newcommand{\non}{\nonumber \\}
\newcommand{\ppa}[2]{\left(\frac{\partial}{\partial #1}\right)^{#2}}
\newcommand{\pp}[2]{\frac{\partial #1}{\partial #2}}
\newcommand{\rn}{Reissner-Nordstr\"om}
\newcommand{\vphi}{\varphi}
\newcommand{\kss}{\left[K_\theta^\theta \right]}
\newcommand{\ktt}{\left[K_\tau^\tau \right]}
\newcommand{\heq}{{\hat =}}
\newcommand{\ecm}{E_{c.m.}}

\title{\bf Non-extremal Kerr black holes as particle accelerators}
\author{ Sijie Gao\footnote{ Email: sijie@bnu.edu.cn} and Changchun Zhong\footnote{Email: cczhong@mail.bnu.edu.cn}\\
Department of Physics, Beijing Normal University,\\
Beijing 100875, China}
\maketitle

\begin{abstract}
It has been shown that extremal Kerr black holes can be used as particle accelerators and  arbitrarily high energy may be obtained near the event horizon. We study particle collisions near the event horizon (outer horizon) and Cauchy horizon (inner horizon) of a non-extremal Kerr black hole. Firstly, we provide a general proof showing that  particles cannot collide with arbitrarily high energies at the outter horizon. Secondly, we show that ultraenergetic collisions can occur near the inner horizon of a Kerr black hole with any spin parameter $a$.

\end{abstract}

\section{Introduction}
Whether Kerr black holes can serve as particle accelerators with infinite collision energy has recently been discussed. Ba\~nados, Silk and West \cite{prl} showed that particles falling from rest outside an extremal Kerr black hole ($a=M$) can collide with  arbitrarily high center of mass energies when the collision occurs arbitrarily close to the horizon. The BSW mechanism has been further discussed and generalized to different spacetimes (See e.g. \cite{obz}-\cite{tom} ).  Jacobson and Sotiriou \cite{jac} pointed out that infinite energies for the colliding particles can only be attained at infinite time. Authors also argued \cite{comment} that such a high energy may not be realizable due to the theoretical upper bound $a/M=0.998$ \cite{thorne}. The BSW mechanism would be more realistic if it worked for non-extremal Kerr black holes ($a<M$). Although numerical analysis indicates that the collision energies are finite near the horizon of a non-extremal Kerr black hole, no rigorous proof has been given. In this paper, we prove analytically that infinite center of mass energies can never be attained outside a non-extremal Kerr black holes. Compared with previous literature, our proof is general in the following senses. Firstly, the collision takes place anywhere outside the black hole, not  confined to the equatorial plane $\theta=\frac{\pi}{2}$. Secondly, we allow the 4-velocities of the two particles to be arbitrary. In particular, the 4-velocities have  non-vanishing $\dot\theta$ components. Finally, we release the restriction that particles fall from infinity, allowing them to fall from anywhere outside the black hole. Our analysis shows that infinite energies can only be attained at the horizon and one of the particles must have a critical angular momentum. However, with such an angular momentum, there always exists a potential barrier outside the black hole preventing the particle from approaching the event horizon.

Since the event horizon of a non-extremal Kerr black hole cannot serve as a particle accelerator creating infinite collision energies, it is natural to ask whether the inner horizon can make a difference. This issue has recently been explored by Lake \cite{lake}. The author claimed in the original version that the center of mass energy for two colliding particles is generically divergent at the inner horizon and no fine tuning is required. Then in the Erratum this claim was withdrawn because physical constraints forbid such collisions. We reexamine this issue in details and arrive at the following conclusions. We first confirm, using different arguments, that a generic divergence at the inner horizon is not possible. We further show that for a critical angular momentum, the center of mass energy diverges at the inner horizon. Such a divergence is similar to that proposed by BSW where a critical angular momentum also is required. The difference is that there is no restriction on the spin parameter $a$. So in principle, infinite collision energies can be obtained near the inner horizon of any non-extremal Kerr black hole.

\section{Collisions near the event horizon $r=r_+$} \label{outer}
In this section, we consider two particles colliding outside a Kerr black hole and show that in any case, they cannot collide with arbitrarily high center of mass energies. The Kerr metric is given by \cite{wald}
\bean
ds^2\eqn-\left(\frac{\Delta-a^2\sin^2\theta}{\Sigma}\right)dt^2-\frac{2a\sin^2\theta(r^2+a^2-\Delta)}{\Sigma}
dtd\phi \non
&+&\left(\frac{(r^2+a^2)^2-\Delta a^2\sin^2\theta}{\Sigma}\right)\sin^2\theta d\phi^2+\frac{\Sigma}{\Delta}dr^2+\Sigma d\theta^2\,,
\eean
where
\bean
\Sigma\eqn r^2+a^2\cos^2\theta \non
\Delta\eqn r^2+a^2-2Mr\,.
\eean
Without loss of generality, we shall choose
\bean
M=1
\eean
in the rest of this paper.
We shall deal with the non-extremal case, i.e.,
\bean
0<a<1  \label{al}\,.
\eean
Suppose a particle of mass $m$ moves in the spacetime with 4-velocity $u^a=\ppa{\tau}{a}$, where $\tau$ is the proper time. The 4-velocity of  one of the particle at the point of collision takes the general form
\bean
u^a=\dot t\ppa{t}{a}+\dot r\ppa{r}{a}+\dot\theta\ppa{\theta}{a}+\dot \phi\ppa{\phi}{a} \label{ua}\,.
\eean
 The geodesic motion is determined by the following conserved quantities \cite{wald}
\bean
E\eqn -g_{ab}u^a\ppa{t}{b}= \left(1-\frac{2r}{\Sigma}\right)\dot t+\frac{2ar\sin^2\theta}{\Sigma}\dot\phi \label{E} \\
L\eqn g_{ab}u^a\ppa{\phi}{b} =-\frac{2ar\sin^2\theta}{\Sigma}\dot t+\frac{(r^2+a^2)^2-\Delta a^2 \sin^2\theta}{\Sigma}\sin^2\theta \dot\phi  \nonumber \\
&& \label{L}\\
-1\eqn g_{ab}u^au^b \label{uu}\,,
\eean
where $E$ is the conserved energy per unit mass and $L$ is the angular momentum per unit mass. Solving \eqs{E} and \meq{L} yields
\bean
\dot t\eqn\frac{a^4E-4aLr+2Er^4+a^2Er(2+3r)+a^2E\Delta\cos2\theta}{2\Delta\Sigma} \label{td}\,,\\
\dot\phi\eqn\frac{a(2Er-aL)+L\Delta\csc^2\theta}{\Delta\Sigma} \label{fd}\,.
\eean
Since $u^a$ is a future-directed timelike vector, it follows that $\dot t>0$ near the horizon $r=r_+$ \cite{wald}. By expanding the numerator of \eq{td} around $r=r_+$ and requiring the leading term to be non-negative, we find
\bean
L\leq\frac{2E}{a}\left(1+\sqrt{1-a^2}\right)\,. \label{lfc}
\eean
 Note that \eq{lfc} was also derived in \cite{overspin} from the null energy condition.

Then \eq{uu} yields
\bean
\dot r=-\frac{1}{2\sqrt 2\Sigma}\sqrt S\label{rds}\,,
\eean
where
\bean
S&=&8\Delta\Sigma(-1-\dot\theta^2\Sigma)-\csc^2\theta\left[-a^4E^2+4a^2L^2-6a^2E^2r+16aELr \right. \non
&-&16L^2r-5a^2E^2r+8L^2r^2-4E^2r^4+a^2E^2\Delta\cos4\theta \non
&+&\left. 4(-4aELr+E^2r^4+a^2L^2+a^2E^2r(2+r))\cos2\theta\right] \label{S}\,.
\eean
Note that we have chosen the minus sign for $\dot r$, referring to ingoing geodesics. We shall discuss the plus sign at the end of this section.

Obviously, a physically allowed trajectory satisfies
\bean
S\geq 0 \label{ssge}\,.
\eean

Suppose two particles with the same mass $m$ collide each other. The center of mass energy is given by \cite{prl}
\bean
\ecm=m\sqrt{2}\sqrt{1-g_{ab}u^au^b_{2}}\,.
\eean
For simplicity, define the effective center of mass energy
\bean
E_{eff}=-g_{ab}u^au^b_{2}\,.
\eean
We find
\bean
E_{eff}=-\Sigma\dot\theta\dot\theta_2-\frac{E'}{8\Delta\Sigma} \label{eef}\,,
\eean
where
\bean
E'\eqn(-a^4 E E_2 + 4a^2L L_2 - 6 a^2EE_2 r +
        8 aE_2Lr + 8aE L_2r - 16 LL_2 r \non
&-&5 a^2 E E_2 r^2 + 8 L L_2 r^2 - 4 E E_2 r^4 +
        4(-2a(E_2 L + EL_2) r + E E2 r^4\non
&+&a^2(L L_2 + E E_2 r(2 + r)))\cos2\theta +
        a^2 EE_2\Delta\cos4\theta)\csc^2\theta\non
&+&\sqrt{S}\sqrt{S_2} \label{epp}\,.
\eean
Here $S_2$ is obtained by replacing $L,E,\dot\theta$ with $L_2,E_2,\dot\theta_2$ in \eq{S}.

Our purpose is to examine whether an infinite $E_{eff}$ defined in \eq{eef} exists under the constraints \meq{al},\meq{lfc}, \meq{ssge}. As physical requiremens, all the constants $E$, $L$, etc. should be finite. Otherwise, infinite collision energies can be produced even in Minkowski spacetime. It is also reasonable to assume that $\dot\theta$ and $\dot\theta_2$ are finite at the horizon. This is because one can define
\bean
L_\theta\equiv g_{ab}u^a\ppa{\theta}{b}=\Sigma \dot\theta \label{ls}
\eean
as the ``angular momentum with respect to $\theta$''. Note that unlike $L$, $L_\theta$ is not constant along geodesics. However, it is plausible to require that $L_\theta$ be finite everywhere, particularly at the horizon. It then follows from \eq{ls} that $\dot\theta$ is finite at the horizon.

Hence, it is easy to see from \eq{eef} that an infinite $E_{eff}$ cannot be obtained unless $\Delta=0$, i.e., the collision must occur at the outer horizon $r=r_+$ or the inner horizon $r=r_-$.
 To check if $E_{eff}$ could be infinite, we expand $E'$ at $r=r_+$ and find
\bean
E'=\alpha_0+\alpha_1(r-r_+)+... \label{epn}\,,
\eean
where
\bean
\alpha_0\eqn8\left[-a^2LL_2+E(-8+4a^2-8\sqrt{1-a^2})E_2+2a(EL_2+E_2L)(1+\sqrt{1-a^2})\right.\non
&+&\left.a^2\sqrt{\left(L-\frac{2E(1+\sqrt{1-a^2})}{a}\right)^2}
\sqrt{\left(L_2-\frac{2E_2(1+\sqrt{1-a^2})}{a}\right)^2}\right]\,.
\eean
Using \eq{lfc}, the square root terms can be simplified and one finds
\bean
\alpha_0=0\,.
\eean
The vanishing of $\alpha_0$ is important because it rules out the divergence of $\ecm$ for generic angular momentums.

\begin{figure}[htmb]
\centering \scalebox{0.5} {\includegraphics{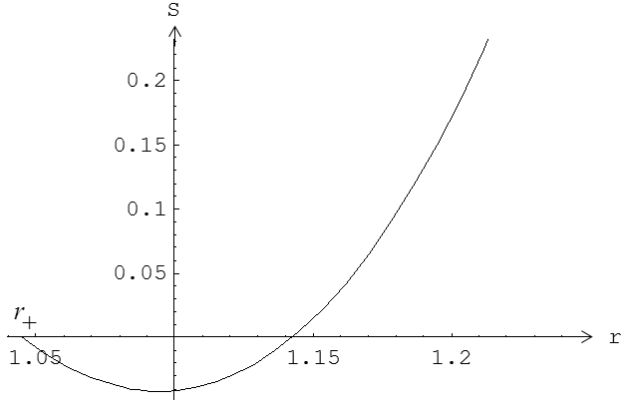}}
\caption{The plot of $S(r)$ in the region $r>r_+$. The parameters are chosen as: $a=0.999$, $\theta=\frac{\pi}{2}$, $L=L_c$. } \label{fig-sr}
\end{figure}

Now it is obvious that $E_{eff}$ cannot be infinite unless $\alpha_1$ is infinite. Since $\sqrt{S}|_{r=r_+}$ appears in the denominator of $\alpha_1$, an infinite $\alpha_1$ requires
\bean
S|_{r=r_+}=2a^2(L-L_c)^2=0\,.
\eean
 Thus, if we choose
\bean
L=L_c=\frac{2E(1+\sqrt{1-a^2})}{a}\,,
\eean
$\ecm$ will become arbitrarily large at the horizon $r=r_+$. However, to make sure that the  particle with this critical angular momentum can actually reach the horizon, \eq{ssge} must hold outside the horizon. By Taylor expansion, we find
\bean
S=a_0+a_1(r-r_+)+...
\eean
For $L=L_c$, $a_0=0$ and $a_1$ is given by.
\bean
a_1=b_1+\left(\frac{2E^2}{a^2\sin^2\theta}+2\dot\theta^2\right)b_2\,,
\eean
where
\bean
b_1\eqn-32+32a^2-32\sqrt{1-a^2}+16 a^2\sqrt{1-a^2}-16a^2\sqrt{1-a^2}\cos^2\theta \nonumber\\
&& \\
b_2\eqn -64+80a^2-16a^4+\sqrt{1-a^2}(-64+48a^2-3a^4-a^4\cos4\theta)\non
&+&4a^2(-4+4a^2-4\sqrt{1-a^2} +4a^2\sqrt{1-a^2})\cos2\theta\,.
\eean
Simple analysis shows that both $b_1$ and $b_2$ are negative for $0<a<1$. Therefore,
\bean
a_1<0 \label{ao}\,,
\eean
 which means $S$ is negative near the horizon and consequently the particle with $L=L_c$ cannot approach the horizon. The barrier outside the horizon is illustrated in \fig{fig-sr}.

In the above argument, we have chosen the minus sign in \eq{rds} for both particles. The same choice was made by BSW. What if the two particles take different signs? In that case, the plus sign in front of $\sqrt S$ in \eq{epp} will become a minus sign and consequently $\alpha_0$ in \eq{epn} will not vanish. Therefore, such two particles will collide with infinite energy even without a fine turning on angular momentum. But to make this happen, one of the particles must be outgoing ($\dot r>0$) on the horizon. For a non-extremal black hole, even holding a particle still at the horizon requires an infinite local force \cite{wald}. Therefore, the configuration with $\dot r>0$ should be ruled out.

\section{Collisions near the inner horizon $r=r_-$.}
\begin{figure}[htmb]
\centering \scalebox{0.6} {\includegraphics{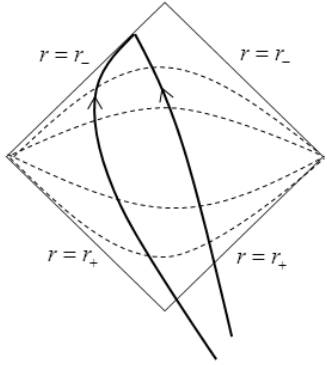}}
\caption{Penrose diagram for the Kerr spacetime in the region $r_-\leq r\leq r_+$. Two particles cross the event horizon and collide at the inner horizon. The dashed lines refer to surfaces of constant $r$.} \label{fig-cauchy}
\end{figure}
Now we discuss the motions in the region $r_-<r<r_+$.  Due to the different natures of the inner and outer horizons, some of the arguments in section \ref{outer} will break down in this section. We shall highlight the differences.

As illustrated in \fig{fig-cauchy}, we consider two particles crossing the event horizon of the right-hand universe and hitting each other at the Cauchy horizon of the left-hand universe. The general equations \meq{ua}-\meq{fd} remain unchanged. However, \eq{lfc} is derived from the fact that $\grad^a t$ is a past-directed timelike vector, which is no longer true outside the inner horizon. Instead, $\grad^a r$  becomes past-directed in this region  and any particle must fall in the direction of $\dot r<0$, i.e.,
\bean
\dot r=-\frac{1}{2\sqrt 2\Sigma}\sqrt S<0 \label{rdm2}\,.
\eean
 It should be noticed that \eq{rds} holds because we have chosen the ingoing mode for both particles in the region $r>r_+$, while \eq{rdm2} holds for any particle in the region $r_-<r<r_+$. Note that \eqs{S}-\meq{epp} remain unchanged. Then we expand $E'$ at $r=r_-$. Corresponding to \eq{epn}, we have
\bean
E'=\alpha_0+\alpha_1(r-r_-)+...
\eean
where
\bean
\alpha_0\eqn8\left[-a^2LL_2+E(-8+4a^2+8\sqrt{1-a^2})E_2+2a(EL_2+E_2L)(1-\sqrt{1-a^2})\right.\non
&+&\left.a^2\sqrt{\left(L-\frac{2E(1-\sqrt{1-a^2})}{a}\right)^2}
\sqrt{\left(L_2-\frac{2E_2(1-\sqrt{1-a^2})}{a}\right)^2}\right]\,.
\eean
This equation can be simplified by choosing the ingoing mode for both particles, as depicted in \fig{fig-cauchy}. Here in the region $r_-<r<r_+$, the ingoing mode means $\dot t<0$ (By the same argument as given at the end of section \ref{outer}, $\dot t>0$ at the left-hand Cauchy horizon is not physically realizable). Then using the same argument which led to \eq{lfc}, we find
\bean
\left(L-\frac{2E(1-\sqrt{1-a^2})}{a}\right)\left(L_2-\frac{2E_2(1-\sqrt{1-a^2})}{a}\right)\geq 0 \,,
\eean
which leads to
\bean
\alpha_0=0\,.
\eean

Again, the necessary condition for  $E_{eff}$ blowing up at the horizon is that $\alpha_1$ blows up, which requires
\bean
S|_{r=r_-}=8a^2(L-L_c')^2=0\,,
\eean
where
\bean
L_c'=\frac{2E(1-\sqrt{1-a^2})}{a}\,.
\eean
Therefore, an infinite collision energy near the inner horizon requires that one of the particles has the momentum
\bean
L=L_c'
\eean
and the other particle has any different angular momentum.  For the collision near the event horizon, we have shown that a potential barrier always exists. But the geodesic motions near the inner horizon are very different. We shall show that there is no potential barrier in the vicinity of the inner horizon. For our purposes, it is sufficient to consider the motions in the equatorial plane and let $E=1$, in which case, $S$ given in \eq{S} reduces to
\bean
S=\frac{16r[a^4-4(1-\sqrt{1-a^2})(r-2)+a^2(r^2+2r+4\sqrt{1-a^2}-8)]}{a^2}\,.
\eean
By Taylor expansion, we find
\bean
S=\frac{32}{a^2}\left[5-a^2-3\sqrt{1-a^2}-\frac{4(1-\sqrt{1-a^2})}{a^2}\right](r-r_-)+...
\eean
By plotting the coefficient of $r-r_-$ as a function of $a$, we see immediately that $S>0$ for any $0<a<1$ in the vicinity of $r=r_-$.

 Now we show, by a concrete example, that such a collision can be realized at the Cauchy horizon for two particles falling from rest at infinity.

The orbits of the two particles are confined to the equatorial plane $\theta=\frac{\pi}{2}$. The parameters are chosen as
\bean
E\eqn E_2=1 \\
L\eqn L_c' \\
L_2\eqn \frac{L'_c}{2}\,.
\eean

\begin{figure}[htmb]
\centering \scalebox{0.4} {\includegraphics{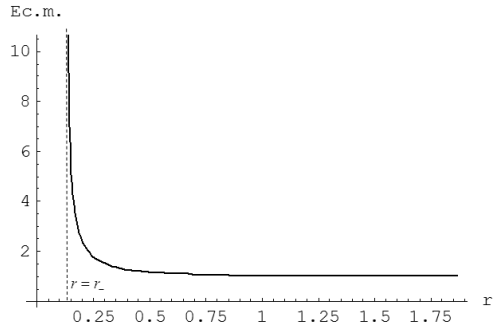}}
\caption{Plot of the center-of-mass energy} \label{fig-ecm}
\end{figure}

\begin{figure}[htmb]
\centering \scalebox{0.6} {\includegraphics{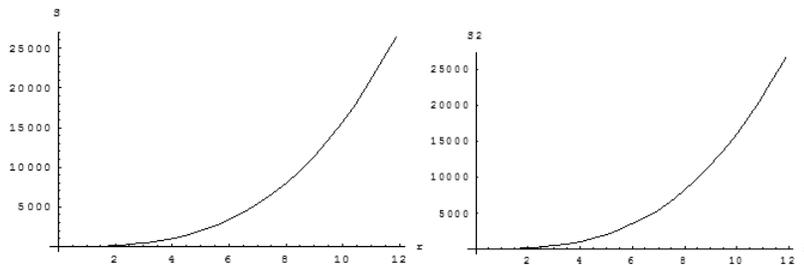}}
\caption{Plots of the functions $S(r)$ and $S_2(r)$. Both functions are positive in the range $r>r_-$. } \label{fig-ss2}
\end{figure}
\begin{figure}[htmb]
\centering \scalebox{0.6} {\includegraphics{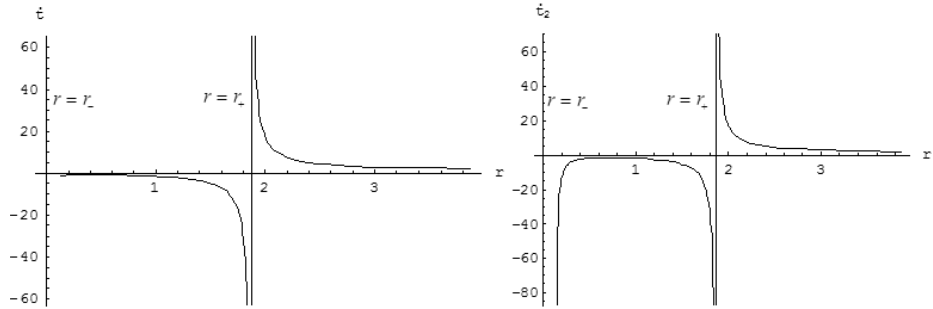}}
\caption{ Plots of $\dot t(r)$ and $\dot t_2(r)$. } \label{fig-td}
\end{figure}

As depicted in \fig{fig-ecm}, the center of mass energy blows up at $r=r_-$. \fig{fig-ss2} shows $S>0$ and $S_2>0$ for all $r>r_-$, i.e., the two particles can fall from infinity all the way to the Cauchy horizon.  \fig{fig-td} plots $\dot t(r)$ in the region $r>r_-$. We see that $\dot t$ and $\dot t_2$ are positive  in the region $r>r_+$, as expected. For $r_-<r<r_+$, $\dot t$ and $\dot t_2$ are negative. This means both particles must hit the left-hand Cauchy horizon, as depicted in \fig{fig-cauchy}. Otherwise, there will be a turning point $\dot t=0$ in this region. It should also be noticed that $\dot t_2\rightarrow -\infty$, meaning that this particle will cross the Cauchy horizon. But for the first particle which has a critical angular momentum, $\dot t$ is finite as $r\rightarrow r_-$. Thus, instead of crossing the Cauchy horizon, this particle spirals asymptotically onto the horizon. This is the necessary mechanism for an infinite center of mass energy as pointed out in \cite{jac}. One can also check that $\dot r$ vanishes on the Cauchy horizon while $\dot r_2$ does not, which is consistent with the behaviors of $\dot t$ and $\dot t_2$.

\section{Conclusions}
We have provided a rigorous proof showing that the center of mass energy cannot be divergent at the event horizon of a non-extremal black hole. This proof is general and exhausts all of the possibilities. The motion of particles is not confined to the equatorial plane and the particles may be released from any point outside the black hole. We find that a critical angular momentum is required for the divergence of $\ecm$ at the horizon. However, with this angular momentum, the particle can never reach the horizon. In relation to Lake's work, we first prove that no divergence of energy occur at the inner horizon for particles with generic angular momentums. A critical angular momentum is required for the divergence. We show with an explicit example that two particles can fall from infinity all the way to the inner horizon and collide with an arbitrarily high center of mass energy.  We have shown that such arbitrarily high energies can be obtained in any non-extremal Kerr spacetime, unlike the case in \cite{prl} that requires the black hole to be exactly extremal. Since extremal black holes do not exist in nature, the ultraenergetic collisions near Cauchy horizons may be more practical and realizable.

\section*{Acknowledgements}
This research was supported by NSFC grants 10605006, 10975016 and
by``the Fundamental Research Funds for the Central Universities''.

\end{document}